\RequirePackage{fix-cm}

\documentclass[twocolumn]{svjour3}
\smartqed
\usepackage{graphicx}
\usepackage{amsmath,amssymb}
\usepackage{mathptmx}
\DeclareMathAlphabet{\mathcal}{OMS}{cmsy}{m}{n}

\usepackage{tabularx,multirow}
\usepackage[pdfstartview=FitH,bookmarks=false,colorlinks=true, linkcolor=blue,citecolor=blue,urlcolor=blue]{hyperref}
\usepackage{cite}
\journalname{Journal of Computational Electronics}

\begin{document}

\title{Two-junction ballistic switch in quantum network model}
\author{	D. E. Tsurikov \and
			A. M. Yafyasov
}
\institute{	D. E. Tsurikov \at
				Institute of Physics, St. Petersburg State University,\\
				Ulyanovskaya 1, St. Petersburg 198504, Russia \\
				\email{DavydTsurikov@mail.ru}\\
				ORCID ID: 0000-0001-6677-7427
			\and
			A. M. Yafyasov \at
				Institute of Physics, St. Petersburg State University,\\
				Ulyanovskaya 1, St. Petersburg 198504, Russia \\
				ORCID ID: 0000-0002-0733-1107 \\
}

\date{\today}

\maketitle

\begin{abstract}
Searching of optimal parameters of nanoelectronic devices is a primal problem in their modeling. We solve this problem on example of the electron ballistic switch in quantum network model. For this purpose, we use a computing scheme in which closed channels are taking into account. It allows calculating correctly a scattering matrix of the switch and, consequently, the electric currents flowing through it. Without losing generality, we consider model of two-junction switch at room temperature. Its character is localization of the controlling electric field in the domain before branching. We optimize switch parameters using a genetic algorithm. At the expense of it for InP, GaAs and GaSb switch efficiency reached 77--78\%. It is established that, for the considered materials, volt-ampere characteristics of the device are close to the linear ones at bias voltages 0--50 mV. It allowed describing with a good accuracy electron transport in the switch by means of $3\times 3$ matrix of approximate conductivity. Finally, based on the performed parameters optimization of two-junction switch we formulate the general scheme of modeling nanoelectronic devices in the framework of quantum network formalism.
\keywords{Ballistic switch \and Quantum network \and Landauer--B\"uttiker formalism \and Extended scattering matrix \and Closed channels \and Genetic algorithm}
\end{abstract}

\section{Introduction}\label{Sec01}

The semiconductor nanoelectronics is a driving factor of development of modern computer equipment. Due to the high level of development of planar technology, devices based on two-dimensional electron gas are especially interesting here. One of perspective two-dimensional nanodevices is the ballistic switch \cite{Bib01,Bib02,Bib03,Bib04,Bib05,Bib06,Bib07,Bib08}. Due to quantum functionality, it has no temperature limitation for the switch voltage \cite{Bib02}. It reduces its power consumption and does more suitable for creation of logical elements of computer equipment \cite{Bib09,Bib10}.

Searching of their optimal parameters is relevant for switches \cite{Bib11}. At the same time, it is necessary to consider correctly features of electron transport in the nanodevice: multichannel scattering, tunneling effect, quantum statistics. The calculation scheme offered recently \cite{Bib12} based on model of quantum network \cite{Bib13,Bib14} allows doing it. In this work we will demonstrate its opportunities for parameters optimization of the switch.

Without losing generality, we will consider two-junction switch~--- the switch in which design the controlling electric field acts in domain before branching. Such design is physically justified owing to the following reasons. Electron transport in the switch is controlled by the external electric field, which usually acts in its branching domain. At the same time, in experiments for three-terminal ballistic junctions it is established that their high-temperature electrical characteristics are qualitatively insensitive to the structure details of the devices \cite{Bib15}. Therefore, the same effect can be expected for the ballistic switch at the room temperature.

\section{Structure and efficiency}\label{Sec02}

\begin{figure}[htb]\center
	\includegraphics{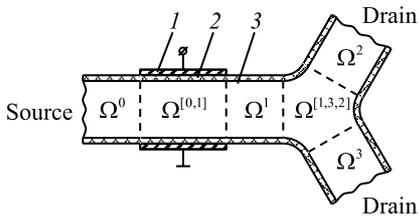}
	\caption{Two-junction switch. \textit{1}~--- metal, \textit{2}~--- dielectric, \textit{3}~--- semiconductor. ${\Omega}^{[0,1]}$ and ${\Omega}^{[1,3,2]}$ are internal junctions, ${\Omega}^1$ is internal branch; ${\Omega}^0$, ${\Omega}^3$ and ${\Omega}^2$ are external branches}\label{Fig01}
\end{figure}

Let us consider the ballistic nanodevice based on two-dimensional electron gas with a gate before Y-junction~--- the \textit{two-junction switch} (fig.~\ref{Fig01}). Let us calculate its transport properties in quantum network model \cite{Bib12}. For this purpose, we divide the switch domain into pieces (dash lines on fig.~\ref{Fig01}) according to their function: \textit{junction} is the network piece in which an electron scatters; \textit{branch} is the network piece in which an electron does not scatter. We consider the device with branches of identical width $B^= $:
\begin{equation}	\label{Eq01}
	\{{B^k} = {B^= }\}^{k\in \mathbb{I}\bigcup \mathbb{E}},
\end{equation}
where $B^k$ is width of $k$th branch, $\mathbb{I} = \{1\}$ and $\mathbb{E} = \{0,3,2\}$ are \textit{tuples} \cite[p.~33]{Bib16} of numbers of internal and external branches respectively (fig.~\ref{Fig01}).

All domains are unambiguously identified based on numbering of network branches (fig.~\ref{Fig01}). Consequently, for the working domain of the switch ${\Omega}^{\langle \text{sw} \rangle}$ we have
\begin{equation}	\label{Eq02}
	{\Omega}^{\langle \text{sw} \rangle} := {\Omega}^{[0,3,2]} = {\Omega}^{[0,1]}\bigcup {\Omega}^1\bigcup {\Omega}^{[1,3,2]},
\end{equation}
where ${{\Omega}^{[0,1]}} = {{\Omega}^{\langle \text{I} \rangle}}$ is I-junction, ${{\Omega}^{[1,3,2]}} = {{\Omega}^{\langle \text{Y} \rangle}}$ is Y-junction. For width of I-junction and joints of Y-junction we have ${B^{\langle \text{I} \rangle}} = {B^{\langle \text{Y} \rangle}} = {B^= }$ and for the arches radius of Y-junction we set ${R^{\langle \text{Y} \rangle}} = {B^= }/2$. External branches of the device ${\Omega}^0$, ${\Omega}^2$ and ${\Omega}^3$ are not included in its working domain as they do not influence its functionality.

In the presence of a potential between a source and drains through the device an electric current flows. The efficiency of the switch is a relative bias of current into one of output branches. Let us choose as desired the branch ${\Omega}^3$. Then the efficiency of the device is formalized as
\begin{equation}	\label{Eq03}
	{\delta}^{\langle \text{sw} \rangle} := {J^3}/({J^2}+{J^3}),
\end{equation}
where $J^k$ is electric current in the external branch ${\Omega}^k$ (Appendix~\ref{AppA}).

\section{Values and ranges of parameters}\label{Sec03}

It is convenient to calculate efficiency of the switch (\ref{Eq03}) in terms of its dimensionless parameters. Let us divide them into two groups: fixed and variable. We will optimize the variable parameters during numerical calculation for achievement of the desired transport properties of the device.

We will find values of fixed and ranges of variable dimensionless parameters based on their dimensional analogs and materials parameters. In this work we consider the switch based on the two-dimensional electron gas formed in III-V semiconductors heterostructures: InP (${m_{\text{e}}} = 0.080{m_0}$), GaAs (${m_{\text{e}}} = 0.063{m_0}$), GaSb (${m_{\text{e}}} = 0.041{m_0}$) \cite{Bib17}. Here $m_{\text{e}}$ is electron effective mass, $m_0$ is mass of free electron.

Taking into account structure of the switch (section~\ref{Sec02}) we set its geometrical characteristics ${B^= } = {B^{\langle \text{I} \rangle}} = {B^{\langle \text{Y} \rangle}} = 10\ \text{nm}$, ${R^{\langle \text{Y} \rangle}} = 5\ \text{nm}$. For length of branch~1 $A^1$ and length of I-junction $A^{\langle \text{I} \rangle}$, we choose the range ${A^1},{A^{\langle \text{I} \rangle}}$ \linebreak $= 10\div 50\ \text{nm}$. Due to such geometrical sizes, size-quantization effects in the switch will be especially essential.

Taking into account the dimensional width of branches $B^= $ and expressions (\ref{Eq35}) and (\ref{Eq14}), we can find dimensional channels energies ${\{E_{\bot n}^= \}}_n$ for the materials relevant in work (tab.~\ref{Tab01}). It is obvious that the distance between energies is more than thermal broadening of levels at room temperature: 0.026 eV. It means that channels energies in system are well resolved.

\begin{table}
\caption{\label{Tab01}Channels energies  ($B^= = 10\ \text{nm}$)}
\centering
\begin{tabularx}{\linewidth}{>{\raggedright}X>{\raggedright}X>{\centering}X>{\centering}X>{\centering}X}
\hline\noalign{\smallskip}
    \multirow{3}*{$n$} &   \multirow{3}*{$\lambda _n^=$}   &\multicolumn{3}{c}{$E_{\bot n}^= $, eV}	\tabularnewline 
\noalign{\smallskip}\cline{3-5}\noalign{\smallskip}
		&			&	InP		&	GaAs	&	GaSb	\tabularnewline 
\noalign{\smallskip}\hline\noalign{\smallskip}
	1	&	9.87	&	0.047	&	0.060	&	0.092	\tabularnewline 
	2	&	39.5	&	0.188	&	0.239	&	0.367	\tabularnewline 
	3	&	88.8	&	0.423	&	0.537	&	0.825	\tabularnewline 
\noalign{\smallskip}\hline
\end{tabularx}
\end{table}

As the range for a Fermi level $E_{\text{F}}^{\langle \text{sw} \rangle}$ of the switch we set the domain between energies of the first and third channel: $E_{\text{F}}^{\langle \text{sw} \rangle}\in (E_{\bot 1}^= ,E_{\bot 3}^= )$. We chose such restriction because in an actual experiment an observation of the highest channels is complicated. External voltages applied in push-pull fashion define Fermi levels in reservoirs (external branches):
\begin{equation}	\label{Eq04}
	E_{\text{F}}^0 = E_{\text{F}}^{\langle \text{sw} \rangle}-{e_0}{U_{\parallel}}/2,\quad E_{\text{F}}^2 = E_{\text{F}}^3 = E_{\text{F}}^{\langle \text{sw} \rangle}+{e_0}{U_{\parallel}}/2,
\end{equation}
where ${U_{\parallel}} := {U^{02}} = {U^{03}}$ is bias voltage (\ref{Eq17}). In problems about electron transport in low-dimensional systems typical values of bias voltage $\sim 10\ \text{mV}$ \cite{Bib15}. We choose for calculations ${U_{\parallel}} = 50\ \text{mV}$.

We suppose that temperatures of all reservoirs are identical (\ref{Eq18}). We consider the switch at room temperature ${T^= } = 300\ \text{K}$. Finally, for an electric field intensity in I-junction we choose range $\mathcal{E}_{\bot}^{\langle \text{I} \rangle} = -{{10}^8}\div 0\ \text{V/m}$.

Based on values and ranges of dimensional parameters, defining quantity
\begin{equation}	\label{Eq05}
	{{\upsilon}_{\parallel}} := 2{m_{\text{e}}}{L^2}{{\hbar}^{-2}}{e_0}{U_{\parallel}},
\end{equation}
taking into account expressions (\ref{Eq28}), (\ref{Eq24}), (\ref{Eq15}), (\ref{Eq05}), (\ref{Eq44}), (\ref{Eq21}) and (\ref{Eq35}) for the optimization procedure we have table~\ref{Tab02}.

\begin{table*}
\caption{\label{Tab02}Values and ranges of dimensionless parameters of two-junction switch ($T^= = 300\ \text{K}$)}
\begin{tabularx}{\linewidth}{lcc*{6}{>{\centering}X}c*{4}{>{\centering}X}}
\hline\noalign{\smallskip}
	\multirow{3}*{SC}	&	\multirow{3}*{${m_{\text{e}}}/{m_0}$}	&	&\multicolumn{6}{c}{fixed parameters}	&	&\multicolumn{4}{c}{variable parameters}	\tabularnewline
\noalign{\smallskip}\cline{4-9}\cline{11-14}\noalign{\smallskip}
	&	&	&	$b^=$	&	$b^{\langle {\rm I} \rangle}$	&	$b^{\langle \text{Y} \rangle}$	&	$r^{\langle \text{Y} \rangle}$	&	${\mu}^=$	&	${\upsilon}_{\parallel}$	&	&	$a^{\text{1}}$	&	$a^{\langle {\rm I} \rangle}$	&	$\epsilon _{\bot}^{\langle {\rm I} \rangle}$	&	$\varepsilon _{\text{F}}^{\langle \text{sw} \rangle}$	\tabularnewline
\noalign{\smallskip}\hline\noalign{\smallskip}
	InP	&	0.080	&	&	1	&	1	&	1	&	0.5	&	5.43	&	10.50	&	&	[1, 5]	&	[1, 5]	&	$[-210, 0]$	&	$[{{\pi}^2},{{(3\pi)}^2}]$	\tabularnewline
	GaAs	&	0.063	&	&	1	&	1	&	1	&	0.5	&	4.27	&	8.27	&	&	[1, 5]	&	[1, 5]	&	$[-165, 0]$	&	$[{{\pi}^2},{{(3\pi)}^2}]$	\tabularnewline
	GaSb	&	0.041	&	&	1	&	1	&	1	&	0.5	&	2.78	&	5.38	&	&	[1, 5]	&	[1, 5]	&	$[-107, 0]$	&	$[{{\pi}^2},{{(3\pi)}^2}]$	\tabularnewline
\noalign{\smallskip}\hline
\end{tabularx}
\end{table*}

\section{Parameters optimization}\label{Sec04}

Let us find a maximum of efficiency (\ref{Eq03}) as function of variable switch parameters $a^1$, $a^{\langle \text{I} \rangle}$, $\epsilon _{\bot}^{\langle \text{I} \rangle}$, $\varepsilon _{\text{F}}^{\langle \text{sw} \rangle}$. For this purpose we optimize them in the given ranges (tab.~\ref{Tab02}), using a computing scheme (Appendix~\ref{AppA}) which takes into account influence of closed channels on transport properties (Appendix~\ref{AppB}). In this case, the scheme takes the following form.

1.	Calculation of Y-junction S-matrix with ${r^{\langle \text{Y} \rangle}} = 1/2$ and ${b^{\langle \text{Y} \rangle}} = 1$ by ND-map method (\ref{Eq37}).

2.	Calculation of I-junction S-matrix with ${b^{\langle \text{I} \rangle}} = 1$ and current values of $a^{\langle \text{I} \rangle}$ and $\epsilon _{\bot}^{\langle \text{I} \rangle}$ according to expressions (\ref{Eq39})--(\ref{Eq43}).

3.	Calculation of switch S-matrix according to expression (\ref{Eq34}) with ${b^1} = 1$ and current value of $a^1$.

4.	Calculation of the dimensionless electric currents based on switch S-matrix according to expressions (\ref{Eq10}) and (\ref{Eq11}) with fixed ${\mu}^= $, ${\upsilon}_{\parallel}$ (tab.~\ref{Tab02}) and current value of $\varepsilon _{\text{F}}^{\langle \text{sw} \rangle}$.

For parameters optimization we organize an operation loop of items 2--4 based on \textit{genetic algorithm} \cite{Bib18}. The loop ends when the global maximum of function ${\delta}^{\langle \text{sw} \rangle}$ is found. We write optimization results in terms of dimensional parameters of the device (tab.~\ref{Tab03}).

\begin{table}
\caption{\label{Tab03}Optimal dimensional parameters and efficiency of two-junction switch ($T^= = 300\ \text{K}$)}
\begin{tabularx}{\linewidth}{lcc*{2}{>{\centering}X}c}
\hline\noalign{\smallskip}
    SC	&   $A^1$, nm   &   $A^{\langle {\rm I} \rangle}$, nm  &   $\mathcal{E}_{\bot}^{\langle {\rm I} \rangle}$, V/m    &   $E_{\text{F}}^{\langle \text{sw} \rangle}$, eV	&   ${\delta}^{\langle \text{sw} \rangle}$  \tabularnewline
\noalign{\smallskip}\hline\noalign{\smallskip}
    InP	&   0   &   13.5    &   $-9.13\cdot {{10}^7}$   &   0.083	&   0.775   \tabularnewline
    GaAs	&   0   &   12.0    &   $-9.06\cdot {{10}^7}$   &   0.120	&   0.773   \tabularnewline
    GaSb	&   0   &   13.5    &   $-9.86\cdot {{10}^7}$   &   0.284	&   0.780   \tabularnewline
\noalign{\smallskip}\hline
\end{tabularx}
\end{table}

Parameters optimization with the alternative bias of current (to the branch~2) gave smaller values of efficiency. Therefore we have chosen the branch~3 as desired in definition of switch efficiency (\ref{Eq03}).

\section{Results interpretation}\label{Sec05}

Let us discuss results of switch parameters optimization (tab.~\ref{Tab03}). As during optimization the length of branch~1 tended to zero, we have set ${A^1} = 0$ in table~\ref{Tab03}. I-junction length for all semiconductors is close to minimum possible value, and the module of electric field intensity is close to maximum one. With decrease of effective mass of electrons, the optimal Fermi level approaches energy of the second channel (tab.~\ref{Tab01}). For all three materials, the relative bias of current to the branch~3 is 77--78\% that speaks about high efficiency of the switch.

\begin{figure*}[htb]\center
	\includegraphics[trim=0 -10 0 0,clip]{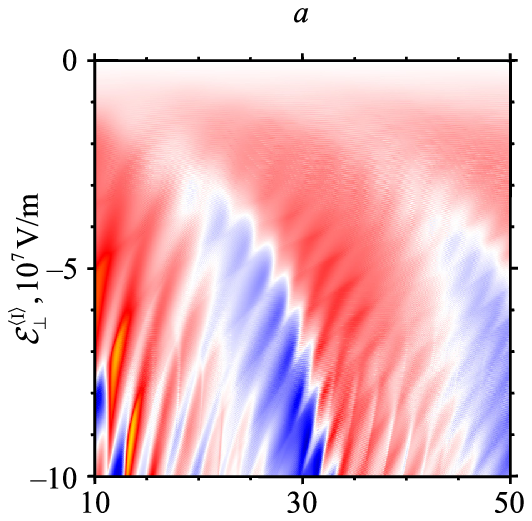} \hfill
	\includegraphics[trim=0 -10 0 0,clip]{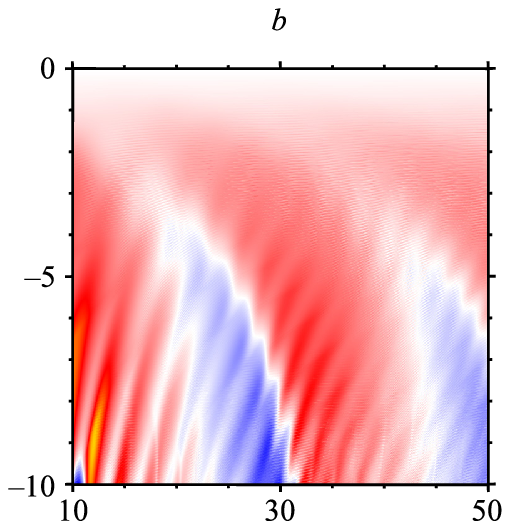} \hfill
	\includegraphics{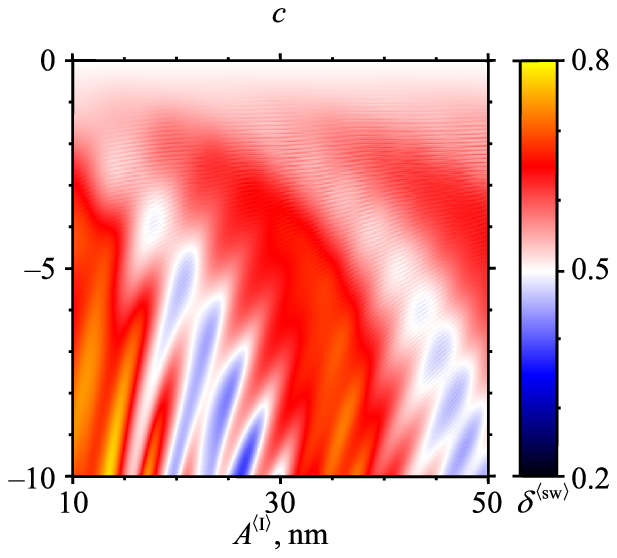}
	\caption{Efficiency (\ref{Eq03}) of two-junction switch (fig.~\ref{Fig01}) at optimal length $A^1$ and Fermi level $E_{\text{F}}^{\langle \text{sw} \rangle}$ (tab.~\ref{Tab03}): \textit{a}~--- InP, \textit{b}~--- GaAs, \textit{c}~--- GaSb}\label{Fig02}
\end{figure*}

Let us fix the optimal length $A^1$ and the Fermi level $E_{\text{F}}^{\langle \text{sw} \rangle}$ for all three materials (tab.~\ref{Tab03}) and plot the graphs of efficiency (\ref{Eq03}) as functions of length $A^{\langle \text{I} \rangle}$ and intensity $\mathcal{E}_{\bot}^{\langle \text{I} \rangle}$ (fig.~\ref{Fig02}). In the figure~\ref{Fig02} we can see domains where bias of current to the branch~3 (red and yellow) and to the branch~2 (blue) dominates and domains without the dominating of bias of current (white). Everywhere these domains replace each other wavy with increase of I-junction length. At the same time, bias of current to the branch~2 begins to be shown only at $\mathcal{E}_{\bot}^{\langle \text{I} \rangle}\approx -5\cdot {{10}^7}\ \text{V/m}$ and less. At $\mathcal{E}_{\bot}^{\langle \text{I} \rangle}>-{{10}^7}\ \text{V/m}$ the dominating bias of current is absent.

In the figure~\ref{Fig02} there are no domains with bias of current in the branch~2 close to the maximal bias of current in the branch~3 (there are no black domains). Calculations showed that this effect is observed in all studied ranges of parameters from table~\ref{Tab02}. From the figure~\ref{Fig02} we can see that it becomes stronger with decrease of electron effective mass, at the same time, domains with bias of current in the branch~3 begin to dominate.

In figures of efficiency for all three materials, there is a large number of local extremums. This phenomenon complicates searching of global maximum of function ${\delta}^{\langle \text{sw} \rangle}$ by means of the algorithms based on gradient-search method. Therefore we executed optimization of switch efficiency in subsection~\ref{Sec04} by means of genetic algorithm \cite{Bib18} that is steady against "rolling" to local extremums.

\begin{figure}[htb]\center
	\includegraphics{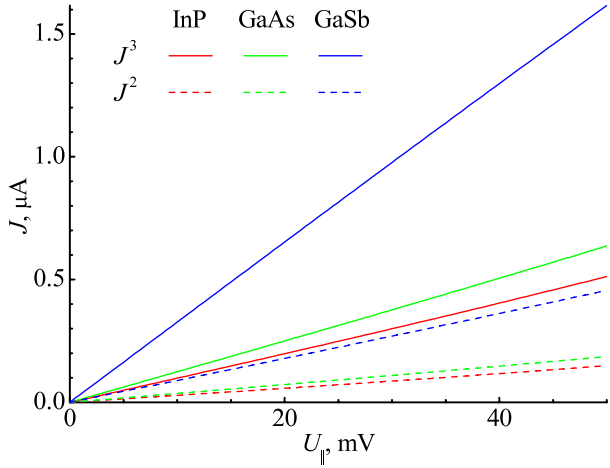}
	\caption{Volt-ampere characteristics of two-junction switch (fig.~\ref{Fig01}) with optimal parameters (tab.~\ref{Tab03})}\label{Fig03}
\end{figure}

Let us construct the volt-ampere characteristics (VAC) of the two-junction switch with optimal parameters from table~\ref{Tab03} (fig.~\ref{Fig03}). For all semiconductors VAC are similar in a form, but differ in value of current. Larger currents are observed through switches based on materials with smaller electrons effective mass. This result is in consent with the experiments, which showed that quantum effects in electric characteristics of three-terminal ballistic junctions based on semiconductors with lower mobility of electrons are expressed more weakly \cite{Bib19}.

Let us supplement VAC of the switch (fig.~\ref{Fig03}) with the table of the relative currents (tab.~\ref{Tab04}). We can see from it that in the considered voltage range the optimal efficiency of the device (\ref{Eq03}) is not less than 77\% for all three materials. The error of the executed calculations we can estimate from below based on of the total current (\ref{Eq16}) in discrepancy value. In this case, it is less than 0.001\% that is acceptable value.

\begin{table*}
\caption{\label{Tab04}Relative currents through two-junction switch (fig.~\ref{Fig01}) with optimal parameters (tab.~\ref{Tab03}) in range ${U_{\parallel}} = 0\div 50\ \text{mV}$}
\begin{tabularx}{\linewidth}{lc*{3}{>{\centering}X}}
\hline\noalign{\smallskip}
	quantity	&	definition	&	InP	&	GaAs	&	GaSb	\tabularnewline
\noalign{\smallskip}\hline\noalign{\smallskip}
	efficiency	&	${J^3}/({J^3}+{J^2})$	&	$>0.774$	&	$>0.\text{773}$	&	$>0.779$	\tabularnewline
	discrepancy	&	$2|({J^0}+{J^3}+{J^2})/({J^0}-{J^3}-{J^2})|$	&	$<8.42\cdot {{10}^{-6}}$	&	$<4.37\cdot {{10}^{-6}}$	&	$<6.21\cdot {{10}^{-6}}$	\tabularnewline
\noalign{\smallskip}\hline
\end{tabularx}
\end{table*}

As VAC of the switch are close to the linear ones (fig.~\ref{Fig03}), its conductivity in considered interval weakly depends on bias voltage. Let us apply to it an approximation of small bias voltages (\ref{Eq20}): ${{\sigma}^{[0,3,2]}}\approx {{\tilde{\sigma}}^{[0,3,2]}}$. At the same time, instead of the type identifier (superscript in angular brackets) we use structural identifier (superscript in square brackets) because according to agreements in work \cite{Bib12} it unambiguously defines structure of matrix:
\begin{equation}	\label{Eq06}
	{{\sigma}^{\langle \text{sw} \rangle}} = {{\sigma}^{[0,3,2]}} = \left[ \begin{matrix}
	{{\sigma}^{[0,3,2]00}} & {{\sigma}^{[0,3,2]03}} & {{\sigma}^{[0,3,2]02}} \\
		{{\sigma}^{[0,3,2]30}} & {{\sigma}^{[0,3,2]33}} & {{\sigma}^{[0,3,2]32}} \\
		{{\sigma}^{[0,3,2]20}} & {{\sigma}^{[0,3,2]23}} & {{\sigma}^{[0,3,2]22}} \\
	\end{matrix} \right].
\end{equation}
For the considered materials we have
\begin{equation}	\label{Eq07}
\begin{aligned}
	\text{InP}:&&\qquad {{\tilde{\sigma}}^{[0,3,2]}} = \left[ \begin{matrix}
		50.2 & \ 9.83 & \ 2.83 \\
		9.83 & \ 12.5 & \ 40.4 \\
		2.83 & \ 40.4 & \ 19.5 \\
	\end{matrix} \right]\ \text{}\!\!\mu\!\!\text{}\,\text{S};
\end{aligned}
\end{equation}
\begin{equation}	\label{Eq08}
	\text{GaAs}:\qquad {{\tilde{\sigma}}^{[0,3,2]}} = \left[ \begin{matrix}
		55.2 & \ 12.4 & \ 3.60 \\
		12.4 & \ 15.9 & \ 42.9 \\
		3.60 & \ 42.9 & \ 24.8 \\
	\end{matrix} \right]\ \text{}\!\!\mu\!\!\text{}\,\text{S};
\end{equation}
\begin{equation}	\label{Eq09}
	\text{GaSb}:\qquad {{\tilde{\sigma}}^{[0,3,2]}} = \left[ \begin{matrix}
		38.8 & \ 32.7 & \ 8.91 \\
		32.7 & \ 21.4 & \ 26.4 \\
		8.91 & \ 26.4 & \ 45.2 \\
	\end{matrix} \right]\ \text{}\!\!\mu\!\!\text{}\,\text{S}.
\end{equation}
From expressions (\ref{Eq07})--(\ref{Eq09}) we see that elements ${\tilde{\sigma}}^{[0,3,2]03}$ and ${\tilde{\sigma}}^{[0,3,2]02}$ grow with decrease of electrons effective mass that is in accord with VAC (fig.~\ref{Fig03}). Conductivities (\ref{Eq07})--(\ref{Eq09}) completely describe quantum electron transport through the switch (fig.~\ref{Fig01}) with optimal parameters (tab.~\ref{Tab03}) at small bias voltages.

\section{Conclusion}\label{Sec06}

In this work, we considered a problem of parameters optimization of the electron ballistic switch in quantum network model. For this purpose, we used a computing scheme in which closed channels are taken into account. It allowed calculating correctly a scattering matrix of the switch and, consequently, the electric currents flowing through it.

Without losing generality, we considered model of two-junction switch at room temperature. Its character is localization of the controlling electric field in the domain before branching. Due to parameters optimization of the switch for InP, GaAs and GaSb its efficiency reached 77--78\% at the sizes $\sim 10\ \text{nm}$ and the intensity module of controlling electric field $<{{10}^8}\ \text{V/m}$. We ploted the graphs of efficiency at optimal length of connecting branch and Fermi level. Multiple local extremums in figures confirmed expediency of the genetic algorithm used during optimization that is steady against "rolling" in them.

We established that, for the considered materials, volt-ampere characteristics of the device are close to the linear ones at bias voltages 0--50 mV. It allowed describing with a good accuracy electron transport in the switch by means of $3\times 3$ matrix of approximate conductivity.

Finally, based on the performed parameters optimization of two-junction switch we can formulate the following scheme of modeling nanoelectronic devices in the framework of quantum network formalism.

1.	\textit{Structure and efficiency}. Selection of device structure according to its function. Formalization of its efficiency in terms of electric currents through it.

2.	\textit{Parameters}. Determination of list, values and ranges of dimensional device parameters. Translation of all quantities to dimensionless ones.

3.	\textit{Optimization}. Searching of local maximum of device efficiency as functions of its variable parameters by means of calculation of dimensionless electric currents through it. Translation of all found quantities to dimensional ones.

4.	\textit{Interpretation}. Analysis of optimization results.

\appendix
\normalsize

\section{Calculation of transport properties of switch}\label{AppA}

In this appendix, we will write the computational method of transport properties of the two-junction switch that we used in sections \ref{Sec04} and \ref{Sec05}.

\paragraph{Electric currents}

Electric currents through the switch we calculate by means of Landauer--B\"uttiker formalism \cite{Bib20,Bib21}. During numerical calculations, it is convenient to write current in an external branch ${\Omega}^k$ in the dimensionless form \cite[(224)]{Bib12}:
\begin{equation}	\label{Eq10}
	\begin{aligned}
		\!{{{\rm I}}^k} = \sum\nolimits_{mn}^l \int_{-\infty}^{+\infty} d\varepsilon  [\lambda _m^k < \varepsilon] &{{| C_{mn}^{\langle \text{sw} \rangle kl}( \varepsilon ) |}^2}[\varepsilon >\lambda _n^l] \\
		\times \{  F_{-1} ([ &\varepsilon _{\text{F}}^l-\varepsilon ]/{{\mu}^l} ) \\
		-F_{-1}  &( [\varepsilon _{\text{F}}^k-\varepsilon ]/{{\mu}^k} ) \},
	\end{aligned}
\end{equation}
\begin{equation}	\label{Eq11}
	{C^{\langle \text{sw} \rangle}} := {K^{+1/2}}{S^{\langle \text{sw} \rangle}}{K^{-1/2}},
\end{equation}
\begin{equation}	\label{Eq12}
	K_{mn}^{kl} := I_{mn}^{kl}\kappa _m^k,\quad \kappa _m^k := \sqrt{\varepsilon -\lambda _m^k},
\end{equation}
where $\lambda _m^k$ is dimensionless energy of $_m^k$th branch-channel (of $m$th channel in the branch ${\Omega}^k$), $\varepsilon $ is dimensionless energy of electron, $C^{\langle \text{sw} \rangle}$ is \textit{extended current scattering matrix} of the switch \cite{Bib22,Bib23}, $S^{\langle \text{sw} \rangle}$ is \textit{extended scattering matrix} of the switch \cite[p.~155]{Bib22}, ${F_{-1}}\left( \eta \right) = 1/(1+{e^{-\eta}})$ is Fermi--Dirac integral of order $-1$, $\varepsilon _{\text{F}}^k$ is dimensionless Fermi level in $k$th branch. Connection of the dimensionless quantities with dimensional ones is set by expressions
\begin{equation}	\label{Eq13}
	{J^k} = :-\frac{{e_0}\hbar}{2\pi {m_{\text{e}}}{L^2}}{{{\rm I}}^k},
\end{equation}
\begin{equation}	\label{Eq14}
	E_{\bot n}^k = {{\hbar}^2}/(2{m_{\text{e}}}{L^2})\cdot \lambda _n^k,\quad E_{\text{F}}^k = {{\hbar}^2}/(2{m_{\text{e}}}{L^2})\cdot \varepsilon _{\text{F}}^k,
\end{equation}
\begin{equation}	\label{Eq15}
	{{\mu}^k} := 2{m_{\text{e}}}{L^2}{{\hbar}^{-2}}{k_0}{T^k}
\end{equation}
for current, energies and temperature respectively. In expressions (\ref{Eq13})--(\ref{Eq15}), $e_0$ is elementary charge, $\hbar $ is Planck's constant, $m_{\text{e}}$ is electron effective mass, $L$ is reference length for transition to the dimensionless problem (it is possible to choose any of convenience reasons), $k_0$ is Boltzmann constant. The error of the current (\ref{Eq13}), which is calculated by formula (\ref{Eq10}), can be estimated by value of the total current
\begin{equation}	\label{Eq16}
	J := \sum\nolimits_{}^{k\in \mathbb{E}}{{J^k}}.
\end{equation}
Theoretically, it is equal to zero according to charge conservation law.

At small voltages between reservoirs $k$ and $l$ (bias voltages)
\begin{equation}	\label{Eq17}
	{U^{kl}} := -( E_{\text{F}}^k-E_{\text{F}}^l)/{e_0}
\end{equation}
and their identical temperature
\begin{equation}	\label{Eq18}
	\{{{\square}^k} = \ {{\square}^= }|\ \square \ = T,\mu \}^{k\in \mathbb{E}}
\end{equation}
we can write currents as
\begin{equation}	\label{Eq19}
	{J^k}\approx \sum\nolimits_{}^l{{{{\tilde{\sigma}}}^{\langle \text{sw} \rangle kl}}{U^{lk}}}
\end{equation}
in terms of approximate conductivity \cite[(227)]{Bib12} of the switch
\begin{equation}	\label{Eq20}
	\begin{aligned}
		{{\tilde{\sigma}}^{\langle \text{sw} \rangle kl}} := \frac{e_0^2}{\pi \hbar}\sum\nolimits_{mn}^{} &\frac{1}{4{{\mu}^= }} \\
		\times \! \int_{-\infty}^{+\infty} d\varepsilon &[\lambda _m^k<\varepsilon ]{{| C_{mn}^{\langle \text{sw} \rangle kl}\left( \varepsilon \right) |}^2}[\varepsilon >\lambda _n^l] \\
		& \!\! \times {\cosh}^{-2}( [{\varepsilon}_{\text{F}} - \varepsilon ]/[2{{\mu}^= }]),
	\end{aligned}
\end{equation}
\begin{equation}	\label{Eq21}
	\varepsilon _{\text{F}}^{\langle \text{sw} \rangle} = 2{m_{\text{e}}}{{\hbar}^{-2}}{L^2}E_{\text{F}}^{\langle \text{sw} \rangle}.
\end{equation}
Here $\varepsilon _{\text{F}}^{\langle \text{sw} \rangle}$ is dimensionless Fermi level of the switch in the absence of voltages on reservoirs. From expression (\ref{Eq20}), we can see that the approximate conductivity does not depend on bias voltages.

Calculations showed that in this work in the sums (\ref{Eq10}) and (\ref{Eq20}) it is enough to consider only the first 4 channels, and it is possible to take energy of the 5th channel for an upper limit of integration:
\begin{equation}	\label{Eq22}
	\sum\nolimits_{mn}^{}{\int_{-\infty}^{+\infty}{d\varepsilon \{...\}}}\mapsto \sum\nolimits_{m,n = 1,...,4}^{}{\int_{-\infty}^{+\infty}{d\varepsilon [\varepsilon <\lambda _5^= ]\{...\}}}.
\end{equation}
At the same time, a lower limit of integration will have values of energies of the first 4 channels ${\{\lambda _m^= \}}_{m = 1,...,4}$.

\paragraph{Switch S-matrix}

According to expressions (\ref{Eq10}) and (\ref{Eq11}), currents through the switch are calculated based on extended scattering matrix ${S^{\langle \text{sw} \rangle}} = {S^{[0,3,2]}}$ of its working domain ${{\Omega}^{\langle \text{sw} \rangle}} = {{\Omega}^{[0,3,2]}}$ (\ref{Eq02}). We find the matrix $S^{[0,3,2]}$ by means of the dimensionless Schr\"odinger equation
\begin{equation}	\label{Eq23}
	( -\partial _1^2-\partial _2^2+\upsilon )\Psi = \varepsilon \Psi .
\end{equation}
Connection of the dimensionless quantities with dimensional ones is set by expressions
\begin{equation}	\label{Eq24}
	\begin{aligned}
		\upsilon({\bf r}) &:= 2{m_{\text{e}}}{{\hbar}^{-2}}{L^2}V(L{\bf r}),\quad \varepsilon := 2{m_{\text{e}}}{{\hbar}^{-2}}{L^2}E, \\
		\Psi({\bf r}) &:= L\varphi(L{\bf r})
	\end{aligned}
\end{equation}
for potential, energy of an electron and wave function of an electron respectively. Following the work \cite{Bib12}, we write the solution of the equation (\ref{Eq23}) in branches of network ${\{{{\Omega}^k}\}}^{k\in \mathbb{I}\bigcup \mathbb{E}}$ as
\begin{equation}	\label{Eq25}
	{{\psi}^{[\mathbb{A}]k}} = {{\psi}^{[\mathbb{A}]\triangleleft k}}+{{\psi}^{[\mathbb{A}]\triangleright k}},\quad k\in \mathbb{A},
\end{equation}
\begin{equation}	\label{Eq26}
	{{\psi}^{[\mathbb{A}]\triangleleft k}} := \sum\nolimits_m{\psi _m^{[\mathbb{A}]\triangleleft k}},\quad {{\psi}^{[\mathbb{A}]\triangleright k}} := \sum\nolimits_m{\psi _m^{[\mathbb{A}]\triangleright k}},
\end{equation}
\begin{equation}	\label{Eq27}
	\begin{aligned}
		\psi _m^{[\mathbb{A}]\triangleleft k}(x, y) &:= [ \exp (-iKx){c^{[\mathbb{A}]\triangleleft}}]_m^kh_m^k(y), \\
		\psi _m^{[\mathbb{A}]\triangleright k}(x, y) &:= [ \exp (+iKx){c^{[\mathbb{A}]\triangleright}}]_m^kh_m^k(y),
	\end{aligned}
\end{equation}
where $c_m^{[\mathbb{A}]\triangleleft k}$ is amplitude of wave, falling on junction ${\Omega}^{[\mathbb{A}]}$ from $_m^k$th branch-channel, $c_m^{[\mathbb{A}]\triangleright k}$ is amplitude of wave, scattered by junction ${\Omega}^{[\mathbb{A}]}$ to $_m^k$th branch-channel, $\mathbb{A}$ is tuple of numbers of branches adjoining junction (for working domain $\mathbb{A} = \mathbb{E}$). Let us choose width of branch (\ref{Eq01}) as the reference length for transition (\ref{Eq24}) to the dimensionless Schr\"odinger equation (\ref{Eq23}):
\begin{equation}	\label{Eq28}
	L = B^=.
\end{equation}
Then for the dimensionless width of branches $b^= $ we have
\begin{equation}	\label{Eq29}
	{b^= } = 1
\end{equation}
and transversal modes $\{h_m^k = h_m^= \}_{m\in \mathbb{N}}^{k\in \mathbb{I}\bigcup \mathbb{E}}$ and channels energies $\{\lambda _m^k = \lambda _m^= \}_{m\in \mathbb{N}}^{k\in \mathbb{I}\bigcup \mathbb{E}}$ are eigenfunctions and eigenvalues of problem
\begin{equation}	\label{Eq30}
	\left\{ \begin{array}{ll}
		-\partial _y^2h_m^= (y) = \lambda _m^= h_m^= (y) &\quad y\in (0, 1) \\
		h_m^= (y) = 0 &\quad y\in \{0, 1\}
	\end{array} \right. .
\end{equation}
Amplitudes of falling and scattered waves, taking into account all channels, are connected with each other by extended scattering matrix $S^{[\mathbb{A}]}$ of junction ${\Omega}^{[\mathbb{A}]}$:
\begin{equation}	\label{Eq31}
	c_m^{[\mathbb{A}]\triangleright k} = \sum\nolimits_n^l{S_{mn}^{[\mathbb{A}]kl}c_n^{[\mathbb{A}]\triangleleft l}}.
\end{equation}

\paragraph{Calculation of switch S-matrix}

The suggested switch model (fig.~\ref{Fig01}) is \textit{QIY-network} \cite[sect.~3.2.1]{Bib12} with structure
\begin{equation}	\label{Eq32}
	\mathcal{N} = \mathcal{Q}\bigcup \mathcal{I}\bigcup \mathcal{Y},
\end{equation}
where $\mathcal{Q} = \varnothing $, $\mathcal{I} = \{\{0,1\}\}$, $\mathcal{Y} = \{\{1,3,2\}\}$ are tuples of structural identifiers of Q-, I- and Y-junctions respectively. Let us calculate an extended scattering matrix of the switch by means of \textit{network formula} \cite[(167)]{Bib12}. As the working domain of the device consists of two junctions connected by one branch (\ref{Eq02}), the network formula takes a form
\begin{equation}	\label{Eq33}
	{S^{\langle \text{sw} \rangle}} = {S^{[0,3,2]}} = {S^{[0,1]}}\otimes {S^{[1,3,2]}}.
\end{equation}
Consequently
\begin{equation}	\label{Eq34}
	\begin{aligned}
		S&^{[0,3,2]} = \left[ \begin{matrix}
		{S^{[0,1]00}} & {O^{03}} & {O^{02}} \\
		{O^{30}} & {S^{[1,3,2]33}} & {S^{[1,3,2]32}} \\
		{O^{20}} & {S^{[1,3,2]23}} & {S^{[1,3,2]22}}
	\end{matrix} \right]+\left[ \begin{matrix}
		{S^{[0,1]01}} & {O^{01}} \\
		{O^{31}} & {S^{[1,3,2]31}} \\
		{O^{21}} & {S^{[1,3,2]21}}
	\end{matrix} \right] \\
		&\times \left[ \begin{matrix}
	-{S^{[0,1]11}} & {U^{[0,1]11}}\exp ( -i{K^{11}}{A^{11}} ) \\
		{U^{[1,3,2]11}}\exp ( -i{K^{11}}{A^{11}} ) & -{S^{[1,3,2]11}} \\
		\end{matrix} \right]^{-1} \\
		&\times \left[ \begin{matrix}
		{S^{[0,1]10}} & {O^{13}} & {O^{12}} \\
		{O^{10}} & {S^{[1,3,2]13}} & {S^{[1,3,2]12}}
	\end{matrix} \right],
	\end{aligned}
\end{equation}
where $O$ is zero matrix, ${S^{[0,1]}} = {S^{\langle \text{I} \rangle}}$ is extended scattering matrix of I-junction, ${S^{[1,3,2]}} = {S^{\langle \text{Y} \rangle}}$ is extended scattering matrix of Y-junction, ${A^{11}} := {{\{{I_{mn}}{a^1}\}}_{mn}}$, $I$ is unit matrix, $a^1$ is dimensionless length of the branch ${\Omega}^1$. Matrix of wave numbers $K^{11}$ and matrices, in charge of change of coordinate frames, $U^{[0,1]11}$ and $U^{[1,3,2]11}$ one can find from the solution of problem (\ref{Eq30}):
\begin{equation}	\label{Eq35}
	\lambda _n^= = (\pi n)^2,\quad h_n^= (y) = \sqrt{2}\sin (\pi ny).
\end{equation}
Then according to definition (\ref{Eq12}) and expression \cite[(236)]{Bib12} we have
\begin{equation}	\label{Eq36}
	\begin{aligned}
		K_{mn}^{11} &= I_{mn}\sqrt{\varepsilon - (\pi n)^2}, \\
		U_{mn}^{[0,1]11} &= U_{mn}^{[1,3,2]11} = (-1)^{m+1}I_{mn}.
	\end{aligned}
\end{equation}
The sizes of submatrices in expression (\ref{Eq34}) correspond to number of the considered channels. In this work, at operations with matrices we consider 6 channels since their further increase weakly influences the calculated electric characteristics. Therefore the size of all submatrices in expression (\ref{Eq34}) is $6\times 6$: ${S^{[0,1]01}} = {{\{S_{mn}^{[0,1]01}\}}_{m,n = 1,...,6}}$, ${O^{30}} = {{\{O_{mn}^{30}\}}_{m,n = 1,...,6}}$, ${K^{11}} = {{\{K_{mn}^{11}\}}_{m,n = 1,...,6}}$ etc. Everywhere among the considered channels there are also closed channels. It allows calculating correctly all S-matrices and therefore currents through the switch (Appendix~\ref{AppB}).

We find matrices $S^{\langle \text{Y} \rangle}$ and $S^{\langle \text{I} \rangle}$ according to the approach that is offered in work \cite[sect.~3.2.2]{Bib12}. For the matrix $S^{\langle \text{Y} \rangle}$, we use direct numerical calculation by the Neumann-to-Dirichlet map method (ND-map) based on triangulation of Y-junction:
\begin{equation}	\label{Eq37}
	{S^{\langle \text{Y} \rangle}} = {{[{N^{\langle \text{Y} \rangle}}iK-I]}^{-1}}[{N^{\langle \text{Y} \rangle}}iK+I],
\end{equation}
where $N^{\langle \text{Y} \rangle}$ is ND-map operator of Y-junction. At the same, time taking into account structure of the switch (section~\ref{Sec02}) and agreements (\ref{Eq28}), for the dimensionless width of joints and arches radius of Y-junction we use
\begin{equation}	\label{Eq38}
	{b^{\langle \text{Y} \rangle}} = {B^{\langle \text{Y} \rangle}}/L = 1,\quad {r^{\langle \text{Y} \rangle}} = {R^{\langle \text{Y} \rangle}}/L = 1/2.
\end{equation}
For the matrix $S^{\langle \text{I} \rangle}$, we use the expression that is obtained in an explicit form based on scattering boundary conditions:
\begin{equation}	\label{Eq39}
	S^{\langle \text{I} \rangle} = G^{\langle \text{I} \rangle \Diamond}{[iK G^{\langle \text{I} \rangle \Diamond} - {\dot{G}}^{\langle \text{I} \rangle \Diamond}]}^{-1}(0)i2K-I,
\end{equation}
\begin{equation}	\label{Eq40}
	\begin{aligned}
		G_{nm}^{\langle \text{I} \rangle \Diamond}\left( x \right) =& \langle h_n^= | h_m^{\langle \text{I} \rangle} \rangle \\
		&\times \left[ \begin{matrix}
			(-1)^{m+1}g_m^{\langle \text{I} \rangle \Diamond 1}(-x) & (-1)^{m+1}g_m^{\langle \text{I} \rangle \Diamond 2}(-x) \\
			g_m^{\langle \text{I} \rangle \Diamond 1}(x + a^{\langle \text{I} \rangle}) & g_m^{\langle \text{I} \rangle \Diamond 2}(x + a^{\langle \text{I} \rangle}) \\
		\end{matrix} \right],
	\end{aligned}
\end{equation}
\begin{equation}	\label{Eq41}
	\begin{aligned}
		h_m^{\langle \text{I} \rangle}(y) =& \ c_{\bot m}^{\langle \text{I} \rangle 1}\operatorname{Ai}( [\epsilon _{\bot}^{\langle \text{I} \rangle}y-\lambda _m^{\langle \text{I} \rangle}]/[\epsilon _{\bot}^{\langle \text{I} \rangle}]^{2/3} ) \\
		&+ c_{\bot m}^{\langle \text{I} \rangle 2}\operatorname{Bi}( [\epsilon _{\bot}^{\langle \text{I} \rangle}y-\lambda _m^{\langle \text{I} \rangle}]/[\epsilon _{\bot}^{\langle \text{I} \rangle}]^{2/3} ),
	\end{aligned}
\end{equation}
\begin{equation}	\label{Eq42}
	\begin{aligned}
		&\left[ \begin{matrix}
			\operatorname{Ai}( -\lambda _m^{\langle \text{I} \rangle}/[\epsilon _{\bot}^{\langle \text{I} \rangle}]^{2/3} ) & \operatorname{Bi}( -\lambda _m^{\langle \text{I} \rangle}/[\epsilon _{\bot}^{\langle \text{I} \rangle}]^{2/3} ) \\
			\operatorname{Ai}( [\epsilon _{\bot}^{\langle \text{I} \rangle}-\lambda _m^{\langle \text{I} \rangle}]/[\epsilon _{\bot}^{\langle \text{I} \rangle}]^{2/3} ) & \operatorname{Bi}( [\epsilon _{\bot}^{\langle \text{I} \rangle}-\lambda _m^{\langle \text{I} \rangle}]/[\epsilon _{\bot}^{\langle \text{I} \rangle}]^{2/3} ) \\
		\end{matrix}\right] \\
		&\times \left[ \begin{matrix}
			c_{\bot m}^{\langle \text{I} \rangle 1} \\
			c_{\bot m}^{\langle \text{I} \rangle 2} \\
		\end{matrix} \right] = 0,
	\end{aligned}
\end{equation}
\begin{equation}	\label{Eq43}
	\begin{aligned}
		g_m^{\langle \text{I} \rangle \Diamond 1}(x) &= \exp( +i\kappa _m^{\langle \text{I} \rangle}x ), \\
		g_m^{\langle \text{I} \rangle \Diamond 2}(x) &= \exp( -i\kappa _m^{\langle \text{I} \rangle}x ),\quad \kappa _m^{\langle \text{I} \rangle} := \sqrt{\varepsilon -\lambda _m^{\langle \text{I} \rangle}},
	\end{aligned}
\end{equation}
where $\epsilon _{\bot}^{\langle \text{I} \rangle}$ is dimensionless electric field intensity in I-junction.

Taking into account that width of I-junction is equal to width of branches ${B^{\langle \text{I} \rangle}} = {B^= } = L$, for the dimensionless parameters we have
\begin{equation}	\label{Eq44}
	\begin{aligned}
		{a^{\langle \text{I} \rangle}} &= {A^{\langle \text{I} \rangle}}/L,\quad {b^{\langle \text{I} \rangle}} = {B^{\langle \text{I} \rangle}}/L = 1, \\
		\epsilon _{\bot}^{\langle \text{I} \rangle} &:= 2{m_{\text{e}}}{L^3}{e_0}{{\hbar}^{-2}}\mathcal{E}_{\bot}^{\langle \text{I} \rangle},
	\end{aligned}
\end{equation}
where $a^{\langle \text{I} \rangle}$ is dimensionless length of I-junction, $b^{\langle \text{I} \rangle}$ is dimensionless width of I-junction.

\section{Influence of closed channels on currents through switch}\label{AppB}

In sections \ref{Sec04} and \ref{Sec05} for calculation of electron transport in the switch, we used a computing scheme (Appendix~\ref{AppA}) in which closed channels are taken into account. In this appendix, we will explain why closed channels influence electric currents through the switch.

Expressions for electric currents (\ref{Eq10}) and for elements of approximate conductivity matrix (\ref{Eq20}) contain only elements of matrix $C^{\langle \text{sw} \rangle}$ which are responsible for connection between open channels: $\{\lambda _m^k<\varepsilon \}_m^{k\in \mathbb{E}}$. Nevertheless, currents and conductivity are influenced by also closed channels: $\{\lambda _m^k\ge \varepsilon \}_m^{k\in \mathbb{E}}$. It is bound to the fact that they need to be considered for correct calculation of a matrix $C^{\langle \text{sw} \rangle}$.

We find matrix $C^{\langle \text{sw} \rangle}$ via extended scattering matrix ${S^{\langle \text{sw} \rangle}} = {S^{[0,3,2]}}$ (\ref{Eq11}). We calculate matrix $S^{[0,3,2]}$ by means of network formula (\ref{Eq34}). All objects in this formula have the following structure
\begin{equation}	\label{Eq45}
	X^{\square} = \left[ \begin{matrix}
		X_{++}^{\square} & X_{+-}^{\square} \\
		X_{-+}^{\square} & X_{--}^{\square} \\
	\end{matrix} \right],
\end{equation}
where symbol $\square $ is any possible superscript corresponding to the given $X = S,O,U,K,A$. Here we also use accepted in literature notation
\begin{equation}	\label{Eq46}
	\{ f_+ := f_{\mathbb{O}}, \ f_- := f_{{\bar{\mathbb{O}}}} \ | \ f = c^{[\mathbb{A}]\triangleleft}, h, ... \},
\end{equation}
where $\mathbb{O} := \{_m^k|\ \lambda _m^k<\varepsilon \}$ is tuple of numbers of \textit{open branch-channels}, $\bar{\mathbb{O}} := \{_m^k|\ \lambda _m^k\ge \varepsilon \}$ is tuple of numbers of \textit{closed branch-channels} \cite[p.~52]{Bib12}. In expression (\ref{Eq45}), the submatrix $X_{++}^{\square}$ is responsible for connection between open channels, $X_{--}^{\square}$~--- between closed channels, $X_{\pm \mp}^{\square}$~--- between closed and open channels.

All objects in formulas for S-matrices of Y- and I- junctions (\ref{Eq37}) and (\ref{Eq39}) also have structure (\ref{Eq45}) with $X = S,N,G,\dot{G}$. At the same time, it is obvious that $N_{\pm \mp}^{\langle \text{Y} \rangle}\ne O_{\pm \mp}^{\langle \text{Y} \rangle}$, $G_{\pm \mp}^{\langle \text{I} \rangle \Diamond}\left( 0 \right)\ne O_{\pm \mp}^{\langle \text{I} \rangle \Diamond}$, $\dot{G}_{\pm \mp}^{\langle \text{I} \rangle \Diamond}\left( 0 \right)\ne O_{\pm \mp}^{\langle \text{I} \rangle \Diamond}$. Then, taking into account formulas (\ref{Eq37}) and (\ref{Eq39}), we have: $S_{\pm \mp}^{\langle \text{Y} \rangle}\ne O_{\pm \mp}^{\langle \text{Y} \rangle}$, $S_{\pm \mp}^{\langle \text{I} \rangle}\ne O_{\pm \mp}^{\langle \text{I} \rangle}$. Therefore closed channels influence the matrix ${S^{[0,3,2]}} = {S^{\langle \text{sw} \rangle}}$ calculated by a network formula (\ref{Eq34}). Consequently, according to definition (\ref{Eq11}), closed channels influence also the elements of matrix $C^{\langle \text{sw} \rangle}$ which are present at expressions for electric currents (\ref{Eq10}) and elements of a matrix of approximate conductivity (\ref{Eq20}).

So that to find correctly currents through the switch, during their calculations it is necessary to consider closed channels.

\bibliographystyle{spmpsci+}
\bibliography{Article}

\end{document}